\documentclass[%
 reprint,
 amsmath,amssymb,
prb,
]{revtex4-1}

\usepackage{graphicx}% Include figure files
\usepackage{dcolumn}% Align table columns on decimal point
\usepackage{bm}% bold math
\usepackage{cases}
\usepackage{extarrows}

\newcommand{\ket}[1]{| #1 \rangle}
\newcommand{\bra}[1]{\langle #1 |}
\newcommand{\avg}[1]{\langle #1 \rangle}

\newcommand{\con}[1]{{#1}}
\newcommand{\rvec}{\mathbf{r}}

\begin{document}

\preprint{APS/123-QED}

\title{Generalized non-equilibrium vertex correction method in coherent medium theory for quantum transport simulation of disordered nanoelectronics}
\author{Jiawei Yan}
\author{Youqi Ke}
\email{keyq@shanghaitech.edu.cn}
\affiliation{Division of Condensed Matter Physics and Photonic Science, School of Physical Science and Technology, ShanghaiTech University, Shanghai, 201203, China}

\date{\today}

\begin{abstract}
  In realistic nanoelectronics, disordered impurities/defects are inevitable and play important roles in electron transport.  However, due to the lack of effective quantum transport method to do disorder average, the important effects of disorders remain largely un-explored or poorly understood.  Here, we report a generalized non-equilibrium vertex correction method with coherent potential approximation for the non-equilibrium quantum transport simulation of disordered nanoelectronics. In this method, the disorder average of various Green's functions are computed by a generalized coherent potential approximation.
  A generalized non-equilibrium vertex correction algorithm is then developed to calculate disorder average of the product of any two real time single-particle Green's functions.
  We obtain nine non-equilibrium vertex corrections and find they can be solved by a set of simple linear equations. As a result, the averaged non-equilibrium density matrix and various important transport properties, including averaged current, disordered induced current fluctuation and the averaged shot noise, can all be efficiently computed in a unified simple scheme. Moreover, the relationship between the non-equilibrium vertex correction method and the non-equilibrium coherent potential approximation theory is clarified, and we prove the non-equilibrium coherent potential equals the non-equilibrium vertex correction and this equivalence is guaranteed by the Keldysh's formulas.
  In addition, a generalized form of conditionally averaged non-equilibrium Green's function is derived to incorporate with density functional theory to enable first-principles quantum transport simulation. Our approach provides a unified, efficient and self-consistent method for simulating non-equilibrium quantum transport through disordered nanoelectronics.
\end{abstract}

\pacs{
85.35.-p,               %nanoelectronic devices
72.25.Mk,               %spin tarnsport through interfaces
73.63.Rt,               %Nanoscale contacts
75.47.De                %GMR
}

\maketitle
%\tableofcontents

\section{Introduction}\label{sec:Introduction}
Due to experimental imperfections or doping for special functionality, disordered impurities/defects are inevitable in realistic nanoelectronic devices. The unintentional disorders can significantly influence the quantum transport properties of device \cite{PhysRevLett.99.076803,PhysRevB.56.2344,Ohno14081998} and give rise to large device-to-device variability.\cite{735728,PhysRevB.88.085420}
Thus, thorough understanding of the effects of disorders is critically important for both modern device technology and fundamental transport physics. However, for experimental investigation of the disorder effects, it is extremely difficult to precisely control the location and concentration of the disorders, if not absolutely impossible, the disorder induced fluctuations make such investigation even more challenging.
It is therefore of great importance to develop a quantum transport method with correct treatment of the disorder effects so that the nonlinear transport properties of disordered nanoelectronics can be predicted from theoretical simulations.

%---------------------------------------------------------------
However, developing such a quantum transport method faces with the following issues:
(\romannumeral1) the non-equilibrium quantum statistics must be correctly treated since electron transport in current flow is an intrinsically non-equilibrium process;
(\romannumeral2) the strong coupling of transport properties to the atomic, chemical and materials details at nanoscale requires accurate atomic-level simulation without using any empirical parameters;
(\romannumeral3) the absence of translational invariance in disordered devices renders many well established state-of-art computational methods useless;
(\romannumeral4) the theoretical transport properties must be averaged over a large ensemble of disorder configurations;
(\romannumeral5) the disorder induced fluctuation of the property needs to be calculated to tell the device-to-device variability.
Since these issues involve different areas of physics, one must combine different theoretical algorithms together to enable quantum transport simulation of disordered devices.
To solve the first two issues, the present workhorse for simulation of ordered nanoelectronics combines non-equilibrium (NE) Green's function (GF) method \cite{keldysh1965diagram,datta1997electronic,haug2008quantum} with the density functional theory (DFT)\cite{hohenberg1964inhomogeneous,kohn1965self,parr1989density} to account for non-equilibrium statistics from atomistic first principles. (Implementation examples are the Ref.\onlinecite{PhysRevB.63.121104,PhysRevB.63.245407,PhysRevB.71.195422,PhysRevLett.96.166804,:/content/aip/journal/jcp/115/9/10.1063/1.1391253,Xue2002151,PhysRevB.65.165401,PhysRevB.70.085410,Rocha2005,PhysRevLett.95.206805}.).
The remaining three issues are basically related to disorder average of the electronic structure, transport property and property fluctuation.
Therefore, It is naturally desired to address the disorder average problem within the NEGF-DFT framework to realize atomistic simulation of the disordered nanoelectronics.

%%------------------------------------------------
A simple method doing disorder average is by enumerating all the possible disorder configurations in a supercell with the size large enough to represent the disorder for a given concentration. However, the computational cost of this `brute force' method is prohibitively large and thus unfeasible for first-principles NEGF-DFT simulations.
Presently, the most effective method to treat disorder in electronic structure calculation is coherent potential approximation (CPA), \cite{PhysRev.156.809,PhysRev.156.1017} which has seen a wide range applications in materials physics. The main idea of CPA is to self-consistently construct a translational invariant effective medium that features the same GF as the averaged one of disordered system and thus the same physical properties as well.
Currently, CPA calculations are mostly carried out with single-site approximation \cite{PhysRev.175.747,PhysRev.178.1136} which decouples the successive scattering events in the random system. For a long period of time, CPA is only applied to treat equilibrium problems, such as calculating equilibrium electronic structure of bulk materials and interfaces, and equilibrium transport properties in combination with a vertex correction\cite{PhysRev.184.614,PhysRevB.2.1771,PhysRevB.73.144421} that accounts for the effects of multiple impurity scattering\cite{RevModPhys.23.287}. This is all because conventional CPA only provides the equilibrium density matrix by calculating the averaged retarded/advanced GF $\avg{G^{R/A}}$ and the conditionally averaged counterpart $\avg{G^{R/A,Q}}$.

However, at non-equilibrium condition, the non-equilibrium density matrix is given by the averaged and conditionally averaged `lesser' GF, namely $\avg{G^<}$ and $\avg{G^{<,Q}}$.
Recently, in Ref.\onlinecite{PhysRevLett.100.166805}, one of the authors (Y.Ke) and his coworkers developed a CPA based non-equilibrium vertex correction (NVC) method to obtain $\avg{G^{<}}$ and $\avg{G^{<,Q}}$, and combined it with NEGF-DFT quantum transport method to enable first principles simulation of disordered nanoelectronics.
The CPA-NVC provides a non-equilibrium effective medium description of the disordered nanoelectronics, and has achieved considerable success in the simulation of disordered nanoelectronics.\cite{PhysRevLett.100.166805,PhysRevLett.105.236801,PhysRevB.79.155406,PhysRevB.81.045406,PhysRevB.84.014401,:/content/aip/journal/apl/101/9/10.1063/1.4748326,PhysRevLett.109.266803,PhysRevB.85.245436,PhysRevB.87.224412, PhysRevB.88.094421,0953-8984-25-42-425301} In this method, the NVC accounts for both effects of the multiple impurity scattering and the non-equilibrium quantum statistics, which it is named after.
Besides, theoretical efforts have also been spent to avoid the NVC by calculating $\avg{G^{<}}$ directly through an approach called non-equilibrium CPA (NECPA). \cite{PhysRevB.85.235111,PhysRevB.88.205415}
Although the derivations of the two NECPAs in Refs.\onlinecite{PhysRevB.85.235111,PhysRevB.88.205415} and CPA-NVC are very different from each other, the two NECPAs are reported to produce the same results as the CPA-NVC method. However, the relationship between the two NECPAs is not clear according to their original literatures, and the internal connection between NECPA and CPA-NVC needs to be clarified.

%-----------------------------------------------------------
Although progresses made by far have enabled the calculation of non-equilibrium electronic structure and averaged electron current for disordered nanoelectronics, the calculation of current fluctuations is still of great challenge for the present methods.
As we shall see in Sec.\ref{sec:Quantum transport properties of disordered device}, disorder induced current fluctuation and the averaged shot noise both require the disorder average of the quantity $ \avg{G^< C G^<} $, while the CPA-NVC or NECPA developed so far can only average a single $\avg{G^<}$.
Rewriting $\avg{G^< C G^<} = \avg{G^{R} \Sigma^{<} G^{A} C G^{R} \Sigma^{<} G^{A}}$ can tell the high complexity of this quantity in which the disorder average connects four correlated and random GFs. Ref.\onlinecite{zhuang2013conductance} reported a perturbation expansion method to calculate the conductance fluctuation and averaged shot noise, and found the convergence is hard to obtain even with high order terms.
Another Ref.\onlinecite{PhysRevB.88.085420} reported that the product of four GFs involves 256 vertex diagrams and they introduced the dressed vertex and dressed double vertex to reduce the 256 diagrams to 6 calculable but very complex diagrams. Therefore, Developing a simple and efficient method to calculate the transport property fluctuations is desirable.

%------------------------------------------------------------
In this paper, we present a generalized CPA-NVC algorithm for the simulation of disordered nanoelectronics at non-equilibrium state.
We provide a generalized CPA formulation in the single-site approximation to derive various disorder averaged GFs.
Based on this generalized CPA, the generalized NVC algorithm is developed to calculate the averaged product of any two real time single-particle GFs, such as $ \avg{G^< C G^<} $.
We obtain nine generalized NVCs which can be solved by a set of simple linear equations.
With the nine generalized NVCs, the disorder averaged non-equilibrium density matrix and various important transport properties, including averaged current, disorder induced current fluctuation and the averaged shot noise, can all be efficiently computed.
In addition, a generalized form of the conditionally averaged NEGF $\avg{G^{<,Q}}$ is derived for multiple disordered components, beyond the binary case reported in the previous CPA-NVC paper.\cite{PhysRevLett.100.166805}
The self-consistent procedures in combination with NEGF-DFT first-principles simulations is also discussed.
The internal connection between the NECPAs and CPA-NVC is clarified and we show the non-equilibrium coherent potential introduced in NECPA equals the NVC.
The generalized CPA-NVC provides a unified, efficient and self-consistent method for simulation of non-equilibrium electron transport properties of disordered nanoelectronics.  

%------------------------------------------------------------
The rest of the paper is organized as follows. 
In Sec.\ref{sec:Quantum transport properties of disordered device}, for a disordered nanoelectronics, we introduce various disorder averaged non-equilibrium quantum transport properties expressed in terms of the product of two GFs.
Sec.~\ref{sec:A Brief Review on the Green's Function Method} reviews the various type of GFs and their relations, and also introduces the general perturbation expansion technique for these GFs.
Sec.\ref{sec:Generalized Coherent Potential Approximation} describes a generalized coherent medium theory in a single-site approximation to provide a formulation of various averaged GFs.
Sec.\ref{sec:Generalized Vertex Correction} formulates the generalized NVC method and clarifies the connection between NECPA and CPA-NVC.
Sec.\ref{sec:First Principle Calculation} derives the generalized conditionally averaged GFs and describes the first-principles calculation using the generalized CPA-NVC in combination with the NEGF-DFT method. Finally, we conclude in Sec.\ref{sec: Conclusion} and provide additional details in Append.\ref{app:Relations between the Langreth Theorem and the Matrix Representation}, \ref{app:GCPA Calculation Decoupling}, \ref{app:Expressing various real-time Green's functions} and \ref{app:Nine Equations for GNVCs}.

\section{Quantum transport properties of disordered device}\label{sec:Quantum transport properties of disordered device}
\begin{figure}
  \centering
  % Requires \usepackage{graphicx}
  \includegraphics[width=8.4cm]{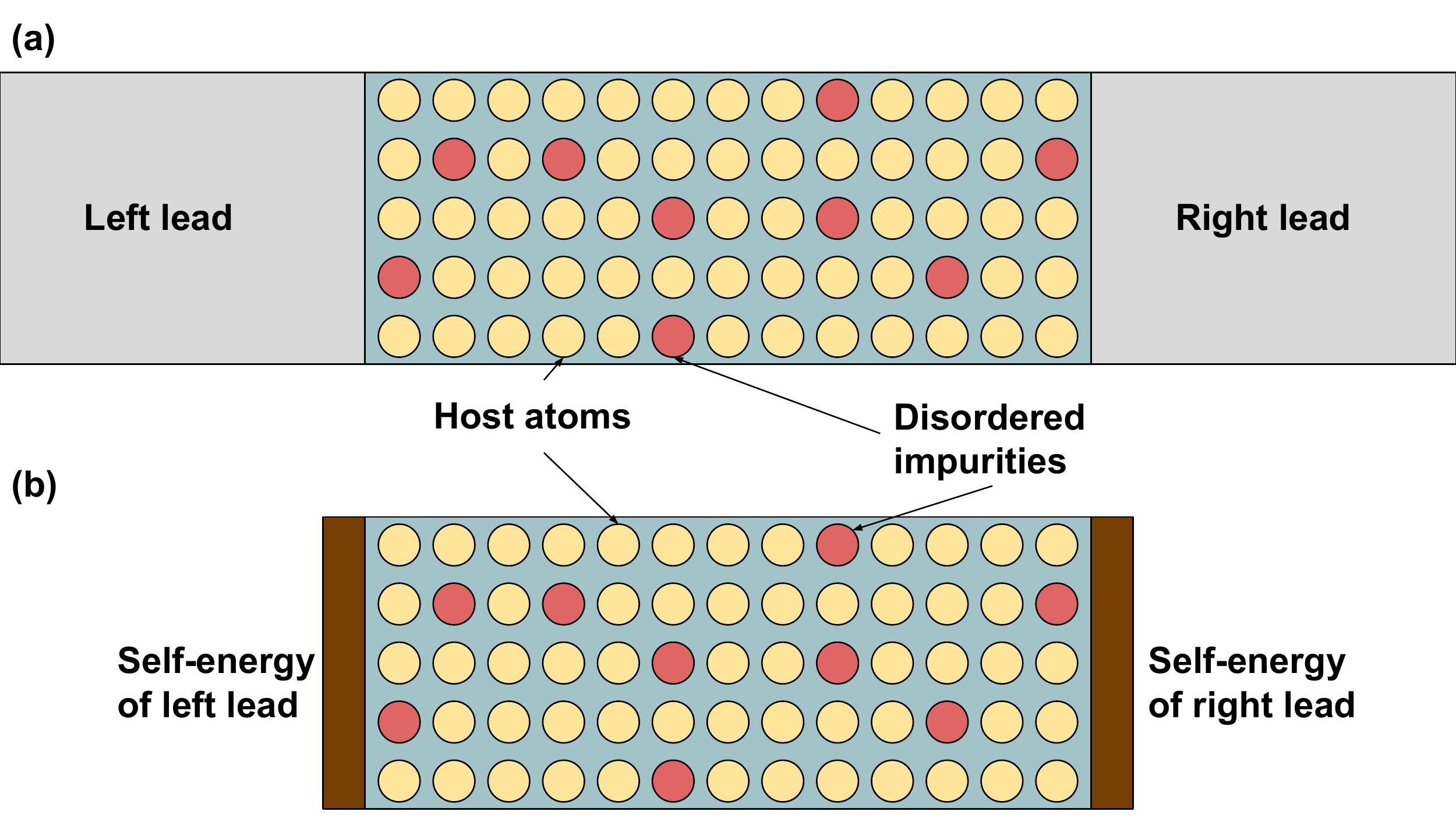}\\
  \caption{(color online) Physical model for a two-probe nanoelectronic device with disorders. (a) The scattering region with disordered impurities sandwiched by two semi-infinite leads. (b) The effects of the leads are turned into the self-energies and changes the infinite and non-periodic system to a finite one.}\label{fig: two-probe system}
\end{figure}

In this section, we briefly review the quantum transport theory based on the NEGF method.
We only consider a two-probe device as shown in Fig.\ref{fig: two-probe system}(a).
The central scattering region containing the disordered impurities is sandwiched by two semi-infinite ideal leads.
Under a finite bias, electrons flow from one lead to the other with scattering events happening on the disordered impurities.
The electron-electron, electron-photon, electron-phonon interactions are not considered in this paper, but in principle, they can be taken into account in the NEGF formalism. \cite{haug2008quantum,mahan2013many}
Since the two-probe device shown in Fig.\ref{fig: two-probe system}(a) is infinite and non-periodic in transport direction, it cannot be calculated directly.
We usually turn the effects of the two semi-infinite leads into the lead self-energies $\Sigma_{ld}$, as shown in Fig.\ref{fig: two-probe system}(b), so that the central device region becomes calculable.

%--------------------------------------------------
For simplicity, we only give some important results for quantum transport properties in the following, and more details can be found in the related literatures.\cite{datta1997electronic,datta2005quantum}
The retarded GF, $G^R$, is directly associated with the Hamiltonian of the central device region $H$ through
\begin{equation}
  G^R = [E - H - \Sigma^R_{ld}]^{-1},
\end{equation}
where $\Sigma^R_{ld} = \Sigma^R_{L} + \Sigma^R_{R}$ is the retarded self-energy due to the left and right leads.
The advanced Green's function and self-energy are conjugate with the retarded ones, namely $G^A = [G^R]^\dag$ and $\Sigma^A_{ld} = [\Sigma^R_{ld}]^\dag$.
Since we assume the leads won't be affected by the scattering region, $\Sigma_{ld}$ is a constant.
For the device shown in Fig.\ref{fig: two-probe system}, the averaged non-equilibrium electron density of the central region is given by
\begin{eqnarray}\label{electron density}
  \avg{\rho(\rvec)} = -i \int \frac{dE}{2\pi} \avg{G^<(\rvec,\rvec';E)}_{\rvec' = \rvec},
\end{eqnarray}
where $\avg{G^<}$ is the averaged lesser Green's function that can be calculated by the Keldysh's formula
\begin{equation}\label{Keldysh formula}
  \avg{G^<} = \avg{G^R \Sigma^<_{ld} G^A}.
\end{equation}
Here, $\Sigma^<_{ld}=\Sigma^<_{L} + \Sigma^<_{R}$ represents the lesser self-energy due to the leads.
It should be mentioned that we have dropped the boundary term in Eq.\eqref{Keldysh formula}, which accounts for the contribution of bound states \cite{PhysRevB.79.045119} and can be neglected for devices in the steady state that we are considering here. \cite{datta1997electronic}
 Since the leads are in equilibrium states, we can get
\begin{equation}\label{Sigma less in terms of Gamma}
  \Sigma_{ld}^< = i [f_{L}(E) \Gamma_{L} + f_{R}(E) \Gamma_{R}],
\end{equation}
where $f_{L/R}(E)$ are the Fermi-Dirac distribution of the left and right leads.
$\Gamma_{L/R}$ in Eq.\eqref{Sigma less in terms of Gamma} are called linewidth functions defined by $\Gamma_{L/R} \equiv i[\Sigma^R_{L/R} - \Sigma^A_{L/R}]$, describing the coupling between the scattering region and the leads.
If we assume $f_L(E) = 1$ and $f_R(E) = 0$, then Eq.\eqref{Sigma less in terms of Gamma} is reduced to
\begin{equation}\label{Sigma less in terms of Gamma reduced}
  \Gamma_L(E) = -i\Sigma_{ld}^<(E),
\end{equation}
and this is what we obtain at zero temperature.
Current through a conductor can be viewed as the probability that electrons travel from one lead to the other.
From Landauer-B\"{u}ttiker formula\cite{datta1997electronic}, the averaged current is given by
\begin{equation}\label{current average}
  \avg{I} = \int\frac{dE}{2\pi}\avg{T(E)}[f_L(E) - f_R(E)],
\end{equation}
where $\avg{T(E)}$ is the averaged transmission coefficient
\begin{equation}\label{transmission coefficient average}
  \avg{T(E)} = Tr\avg{G^R \Gamma_L G^A \Gamma_R}.
\end{equation}
The current fluctuation $\delta I$ under low bias can be approximated by\cite{PhysRevB.88.085420}
\begin{equation}\label{current fluctuation}
  \delta I \approx \int \frac{dE}{2\pi}\delta T(E)[f_L(E) - f_R(E)],
\end{equation}
where the transmission fluctuation is defined as
$\delta T = \sqrt{\avg{T^2} - \avg{T}^2}$,
which involves averaging the square of $T$.
By writing $\avg{T^2}$ explicitly, we have
\begin{equation}\label{T square}
\begin{split}
  \avg{T^2} &= \avg{Tr[G^R \Gamma_L G^A \Gamma_R]\cdot Tr[G^R \Gamma_L G^A \Gamma_R]}\\
            &= -\avg{Tr[G^< \Gamma_R] \cdot Tr[G^< \Gamma_R]},
\end{split}
\end{equation}
where we have used Eq.\eqref{Keldysh formula} and Eq.\eqref{Sigma less in terms of Gamma reduced}.
This equation requires us to average a product of two traces, which is inconvenient in calculation.
To go further, we make a decomposition\cite{PhysRevB.79.045119} $\Gamma_R = \sum_i \ket{W_i}\bra{W_i}$, where $\ket{W_i}$ is the normalized eigenvector of $\Gamma_R$.
By putting this decomposed $\Gamma_R$ into Eq.\eqref{T square} and using the cyclic invariance property of the trace, we get
\begin{equation} \label{T2Glesser}
  \avg{T^2} = -\sum_i \sum_j Tr\avg{G^< S_{ij} G^< S_{ij}^\dag},
\end{equation}
where $S_{ij} \equiv \ket{W_i}\bra{W_j}$ is independent of disorder.
Additionally, the shot noise\cite{PhysRevB.46.12485,PhysRevB.60.16900} given as
\begin{equation}
\begin{split}
\avg{S}=&\int d\epsilon Tr\avg{T}[f_L(1-f_L) + f_R(1-f_R)]\\
&+ \int d\epsilon Tr[\avg{T} - \avg{T^2})](f_L - f_R)^2
\end{split}
\end{equation}
also involves averaging $T^2$ that can be treated in the same way as Eq.\eqref{T2Glesser}.

%------------------------------------------------------------------------------------
By here, we have seen that many physical quantities in electron transport, such as the averaged non-equilibrium electron density, averaged current, current fluctuation and shot noise, can all be expressed in terms of the products of two single-particle GFs. In the following sections, we will discuss how to average these two-GF correlators so that the mentioned quantum transport quantities can be computed for disordered devices.

\section{the Non-equilibrium Green's Function Theory}\label{sec:A Brief Review on the Green's Function Method}
We have introduced several kinds of single-particle GFs to express the non-equilibrium quantum transport properties.
In general, any physical property of single particle operator can be expressed in terms of single-particle Green's functions.
In this section, we will briefly introduce the NEGF theory and the relations between various GFs. Discussion will also cover the perturbation expansion technique which is widely used in NEGF theory.

\subsection{Quantity representation in NEGF theory}
\begin{figure}
  \centering
  % Requires \usepackage{graphicx}
  \includegraphics[width=8.4cm]{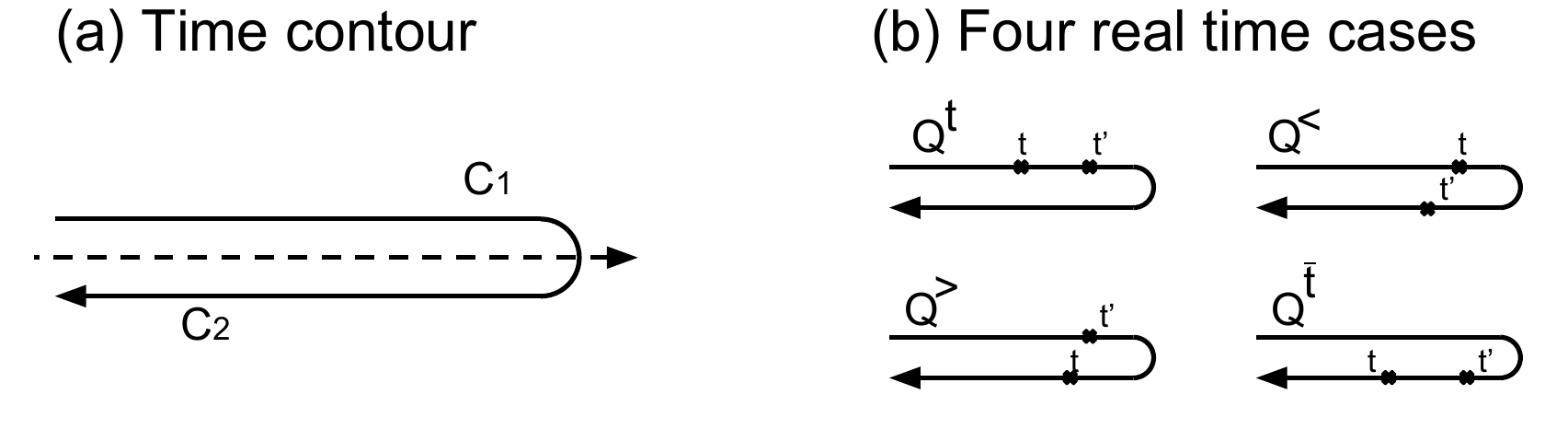}\\
  \caption{(a) Definition of the time-contour, beginning from $-\infty$, passes through $t$ and $t'$, and finally returns to $-\infty$. (b) Four possible combinations of $t$ and $t'$ on the time contour.}\label{fig: contour-loop}
\end{figure}

The central quantity in NEGF theory is the contour-ordered Green's function which is defined as \cite{haug2008quantum}
\begin{equation}\label{contour-ordered Green's function}
  \con{G}(\rvec,t;\rvec',t') \equiv -i\avg{\Phi_0|T_C[ {\psi_H(\rvec,t)\psi_H^\dag(\rvec',t')}]|\Phi_0},
\end{equation}
where $\psi_H$ is the field operator defined in Heisenberg picture, and $\psi_H^\dag$ is its conjugate.
$\ket{\Phi_0}$ refers to the normalized ground state of the system.
$T_C$ is the contour-ordering operator that arranges the time-dependent operators according to their order on the time contour, which starts from remote past, passes through $t$ and $t'$, and finally returns to remote past again, as shown in Fig.\ref{fig: contour-loop}(a).
The reason why the contour looks like this way is because we are considering the non-equilibrium process, in which we can't predict the system when $t\rightarrow +\infty$.

The same as the contour-ordered GF, in NEGF theory, many other physical quantities, such as Halmitonian $H$, self-energy $\Sigma$, potential $V$ and T-matrix $T$ introduced in next section, etc., denoted as $\con{Q}$, can be defined on the time contour.
For these contour-ordered quantities, the time-labels can lie on either of the two branches $C_1$ and $C_2$ on the contour as shown in Fig.\ref{fig: contour-loop}(b). As a result, each contour-ordered quantity contains four different possibilities given as follows
\begin{equation}\label{four real-time quantities}
    \con{Q}(t,t') =
    \begin{cases}
      Q^t(t,t')\qquad &t\in C_1, t'\in C_1,\\
      Q^<(t,t')\qquad &t\in C_1, t'\in C_2,\\
      Q^>(t,t')\qquad &t\in C_2, t'\in C_1,\\
      Q^{\bar{t}}(t,t')\qquad &t\in C_2, t'\in C_2,
    \end{cases}
\end{equation}
which are called time-ordered, lesser, greater and anti-time-ordered real-time quantities, respectively.
One can check that these four real-time quantities are not linearly independent since they satisfy $Q^t + Q^{\bar{t}} = Q^< + Q^>$.\cite{mahan2013many}
Conventionally, we define another three real-time quantities:
\begin{eqnarray}
  Q^R &= Q^t - Q^< &= Q^> - Q^{\bar{t}}\label{retarded quantity},\\
  Q^A &= Q^t - Q^> &= Q^< - Q^{\bar{t}}\label{advanced quantity},\\
  Q^K &= Q^t + Q^{\bar{t}} &= Q^> + Q^<\label{Keldysh quantity}.
\end{eqnarray}
These three terms are called retarded, advanced and Keldysh's quantities and have the relations $Q^R = [Q^A]^\dag$ and $Q^K = -[Q^K]^\dag$.
If $\con{Q}$ is Hermitian, then $Q^K = 0$ and $Q^R = Q^A$.

With the help of these real time quantities, the contour-ordered quantity can be alternatively represented by using a 2-by-2 real-time matrix defined in the following form, as suggested by R.A.Craig,
\begin{equation}\label{Craig 2-by-2 matrix}
  \con{Q} = \left(
              \begin{array}{cc}
                Q^t & -Q^< \\
                Q^> & -Q^{\bar{t}} \\
              \end{array}
            \right),
\end{equation}
which contains the same amount of information as the time-contour representation. %, such as Eq.\eqref{contour-ordered Green's function}.
Since the four elements in above matrix are not linearly independent, to eliminate this redundance, one can apply the Keldysh linear transformation\cite{keldysh1965diagram} $\con{Q}' = R^{-1} \con{Q} R$ to the Craig's matrix, where
$
R \equiv
\frac{1}{\sqrt{2}}
   \begin{tiny}
   \left(
    \begin{array}{cc}
      1 & 1 \\
      -1 & 1 \\
    \end{array}
  \right)
  \end{tiny}
  .
$
After the transformation, we get the Keldysh's 2-by-2 matrix
\begin{equation}\label{Keldysh 2-by-2 matrix}
  \con{Q'} = \left(
              \begin{array}{cc}
                Q^A & 0 \\
                Q^K & Q^R \\
              \end{array}
            \right),
\end{equation}
where the relations in Eqs.\eqref{retarded quantity}-\eqref{Keldysh quantity} are used.
Since the Craig's and Keldysh's representations are related by a simple linear transformation, the real-time Keldysh matrix in Eq.\eqref{Keldysh 2-by-2 matrix} can fully represent a contour-ordered quantity as well.
Note that the Keldysh's matrix features a lower triangular matrix, the matrix addition, multiplication, and inverse operations on  Keldysh's matrices do not change the mathematical structure.
Moreover, the zero element in Eq.\eqref{Keldysh 2-by-2 matrix} can greatly simplify the matrix operations.
For these reasons, it is convenient to work in the Keldysh's representation, and then make the transformation to Craig's matrix to obtain all the real-time GF elements in Eq.\eqref{Craig 2-by-2 matrix}.
As shown in Append.\ref{app:Relations between the Langreth Theorem and the Matrix Representation}, one can assert the equivalence of using different representations of contour-ordered quantity in applications, including Craig's/Keldysh's real-time matrix and time-contour representations. Therefore, we conclude the two NECPAs reported in Ref.\onlinecite{PhysRevB.85.235111,PhysRevB.88.205415} start from the same physical foundation.

\subsection{Perturbation expansion of the Green's function}
After describing the different quantity representations in the NEGF theory, we introduce its central power.
The most appreciate feature of the NEGF theory is its perturbation expansion\cite{haug2008quantum} technique for treating many different kinds of complex interactions, such as interation with electrodes, random impurity scattering and electron-phonon/photon/electron interactions. In particular, an unknown GF of a system can be expanded to an infinite series with definitely calculable quantities. In the presence of some complex interaction, one divides the system Hamiltonian $H$ into two parts
\begin{equation}
  \con{H} = \con{H}_0 + \con{\Sigma},
\end{equation}
where $\con{H}_0$ refers to an unperturbed Hamiltonian that can be calculated exactly, and $\con{\Sigma}$ is the self-energy due to the complex interaction in the system.
A very important result by using the perturbation expansion is that
\begin{equation}\label{perturbation expansion}
  \con{G} = \con{G}_0 + \con{G}_0 \con{\Sigma} \con{G}_0 + \con{G}_0 \con{\Sigma} \con{G}_0 \con{\Sigma} \con{G}_0 + \cdots,
\end{equation}
where GF of realistic system, $\con{G}$, is expanded as an infinite series in terms of $\con{G}_0$ of the unperturbed $H_0$ and $\con{\Sigma}$.
Eq.\eqref{perturbation expansion} can be rewritten in a more compact form, namely the Dyson equation
\begin{equation}\label{Dyson equation}
  \con{G} = \con{G}_0 + \con{G}_0 \con{\Sigma} \con{G} = \con{G}_0 +\con{G} \con{\Sigma} \con{G}_0,
\end{equation}
which is satisfied by many GFs, including the retarded/advanced and different representations of contour ordered GFs.\cite{Craig1968Perturbation,keldysh1965diagram,haug2008quantum}
By replacing quantities with the 2-by-2 real-time matrices, the relations between various real-time GFs in the presence of interaction $ \Sigma $ can be derived from simple matrix multiplication, such as the Keldysh formula\cite{mahan2013many} in Eq.\eqref{Keldysh formula}. Eq.\eqref{Dyson equation} provides a important basis to treat various complex problems in the GF theory, such as the disorder average problem that we want to solve in this paper.

\section{Generalized Coherent Potential Approximation}\label{sec:Generalized Coherent Potential Approximation}

Conventional CPA formulation\cite{PhysRev.156.809,PhysRev.156.1017} of the averaged retarded/advanced GFs $\avg{G^{R/A}}$ are based on the Dyson equation for $G^{R/A}$.
Because of the fact that the contour-ordered GF takes the same form of the Dyson equation as the retarded/advanced GFs,
it is straightforward to extend the conventional CPA formulation to a general case $\avg{\con{G}}$. 
In this section, we introduce a generalized CPA to calculate the disorder average of various GFs introduced above.

\subsection{Theory of generalized CPA}

For the system as shown in Fig.\ref{fig: two-probe system}(b), because of the unintentional impurities, the potential of the system $\con{V}$ is random.
In muffin-tin approximation, $\con{V}$ can be writen as the contribution from each cell centered on atomic nuclear, namely $\con{V} = \sum_n \con{v}_n$, where $\con{v}_n$ is the on-site random potential.
The Hamiltonian of this system can be divided into
\begin{equation}
  \con{H} = \con{H}_0 + \con{\Sigma}_{ld} + \con{V},
\end{equation}
where $\con{H}_0$ is a perfect system Hamiltonian and $\con{\Sigma}_{ld}$ is the self-energy due to the leads.
The central idea of CPA is to construct a coherent effective medium whose GF $\bar{\con{G}}$ is equal to the disorder averaged GF $\avg{G}$ of the system, namely
\begin{equation}\label{central idea of GCPA}
  \bar{\con{G}} = \avg{\con{G}}.
\end{equation}
To physically describe this effective medium, we introduce a self-energy due to disorders $\con{\Sigma}_{im} = \sum_n \con{\Sigma}_{im,n}$, which contains contribution from each site, and rewrite the disordered system Hamiltonian as
\begin{equation} \label{HamilDisord}
  \con{H} = (\con{H}_0 + \con{\Sigma}) + (\con{V} - \con{\Sigma}_{im}),
\end{equation}
where $\con{\Sigma} = \con{\Sigma}_{ld} + \con{\Sigma}_{im}$
contains the contributions from both the leads and effective medium.
The term in the first bracket in Eq.\eqref{HamilDisord} can be regarded as the Hamiltonian of the effective medium and the second bracket contains the deviation of random potential from $\Sigma_{im}$ which can be rewritten as
\begin{equation}\label{on-site relative potential}
  \con{V} - \con{\Sigma}_{im} = \sum_n (\con{v}_n - \con{\Sigma}_{n,im}).
\end{equation}
According to the perturbation expansion technique, we directly write down the Dyson equations for the GFs $\con{G}$ and $\bar{\con{G}}$:
\begin{eqnarray}
  \con{G} &=& \bar{\con{G}} + \bar{\con{G}}(\con{V} - \con{\Sigma}_{im})\con{G} = \bar{\con{G}} + \con{G}(\con{V} - \con{\Sigma}_{im})\bar{\con{G}},\label{Dyson equation for realistic Green's function}\\
  \bar{\con{G}} &=& \con{G}_0 + \con{G}_0 \con{\Sigma} \bar{\con{G}} = \con{G}_0 + \bar{\con{G}} \con{\Sigma} \con{G}_0,\label{Dyson equation for effective medium Green's function}
\end{eqnarray}
where $G$, $\bar{G}$ and $G_0$ are the GFs corresponding to the Hamiltonians $H$, $H_0+\Sigma$ and $H_0$, respectively.
Here, Eq.\eqref{Dyson equation for realistic Green's function} can be rewritten in another form
\begin{equation}\label{T matrix equation}
  \con{G} = \bar{\con{G}} + \bar{\con{G}}\con{T}\bar{\con{G}},
\end{equation}
where $\con{T}$ is called T-matrix defined as
\begin{equation}\label{t-matrix expansion}
\begin{split}
  \con{T} &\equiv (\con{V} - \con{\Sigma}_{im}) + (\con{V} - \con{\Sigma}_{im})\bar{\con{G}} (\con{V} - \con{\Sigma}_{im}) +  \cdots\\
          &= (\con{V} - \con{\Sigma}_{im})(\con{I} + \bar{\con{G}}\con{T}) = (\con{I} + \con{T}\bar{\con{G}})(\con{V} - \con{\Sigma}_{im}).
\end{split}
\end{equation}
From Eq.\eqref{T matrix equation}, we can see T-matrix contains all the complexities of a disordered system. By taking average on both sides of Eq.\eqref{T matrix equation} and comparing with Eq.\eqref{central idea of GCPA}, we obtain an important equation for T-matrix,
\begin{equation}\label{T equals zero}
  \avg{\con{T}} = 0.
\end{equation}

In principle, the above equation in combination with Eq.\eqref{Dyson equation for effective medium Green's function} provides a closed set of self-consistent equations to solve $\Sigma_{im}$ and $\bar{G}$ of the effective medium.
However, to evaluate $\avg{T}$ in Eq.\eqref{T equals zero}, one needs to enumerate all possible configurations of the disorders, which is computational prohibitive.  Therefore, further approximation to the average of T-matrix is required to enable CPA self-consistent calculation.

\subsection{Single-site approximation}

In order to make Eq.\eqref{T equals zero} practically useful, single-site approximation (SSA)\cite{PhysRev.175.747} was introduced to decouple all the disorder scattering events contained in the $T$.
To formulate SSA,
we insert Eq.\eqref{on-site relative potential} into Eq.\eqref{t-matrix expansion} and get
\begin{equation}\label{T in terms of Q}
  \con{T} = \sum_n (\con{v}_n - \con{\Sigma}_{n,im})(\con{I} + \bar{\con{G}} \con{T}) \equiv \sum_n \con{Q}_n,
\end{equation}
where $\con{Q}_n \equiv (\con{v}_n - \con{\Sigma}_{n,im})(\con{I} + \bar{\con{G}} \con{T})$ can be solved to obtain
\begin{eqnarray}
  \con{Q}_n &=& \con{t}_n (\con{I} + \bar{\con{G}} \sum_{m\neq n} \con{Q}_m),\label{Q iterative}\\
  \con{t}_n &\equiv& [I-(\con{v}_n-\Sigma_{n,im})\bar{\con{G}}]^{-1}(\con{v}_n-\Sigma_{n,im}).\label{definetn}
\end{eqnarray}
Here, $\con{t}_n$ describes the scattering event on the single site $n$ ($\con{t}_n=0$ at the site without random occupations).
By recursively substituting Eq.\eqref{Q iterative} into Eq.\eqref{T in terms of Q}, we get the multiple scattering equation:
\begin{equation}\label{multiple scattering equation for T}
  \con{T} = \sum_n \con{t}_n + \sum_{n \neq m}\sum_m \con{t}_n \bar{\con{G}} \con{t}_m + \cdots.
\end{equation}
From this equation, we can see that the overall disorder scattering effects during electron transport are regarded as successive multiple scattering processes from one site to another.
For example, the first two terms in Eq.\eqref{multiple scattering equation for T} are contributed by the respective one-time and two-time scattering processes.
From Eq.\eqref{multiple scattering equation for T}, we can also see the process that an electron is successively scattered twice on a same site is prohibited.
Averaging Eq.\eqref{multiple scattering equation for T} gives
\begin{equation}\label{averaged T before SSA}
  \avg{\con{T}} = \sum_n \avg{\con{t}_n} + \sum_{n \neq m} \avg{\con{t}_n \bar{\con{G}} \con{t}_m} + \cdots.
\end{equation}
By here, all the formulations are exact.
To introduce the SSA, we take the disorder average on Eq.\eqref{Q iterative} and rewrite it as
\begin{equation*}
\begin{split}
  \avg{\con{Q}_n}
  &= \avg{\con{t}_n}(\con{I} + \bar{\con{G}} \sum_{m\neq n} \avg{\con{Q}_m}) + \avg{\con{t}_n \bar{\con{G}} \sum_{m\neq n}(\con{Q}_m - \avg{\con{Q}_m})},
\end{split}
\end{equation*}
where the first term describes the averaged wave scattered by the individual atom on site R, and the second term contains fluctuations away from the average wave.
Neglecting the second term yields the single-site approximation, namely
\begin{equation}\label{averaged Q_n after SSA}
  \avg{\con{Q}_n} = \avg{\con{t}_n}(\con{I} + \bar{\con{G}} \sum_{m\neq n}\avg{\con{Q}_m}),
\end{equation}
which means the successive scattering events are independent of each other.
Since the probability is small for scattering off multiple impurities at the same time, the SSA is a good
approximation and becomes accurate at low impurity concentration.

After applying SSA, we can rewrite Eq.\eqref{averaged T before SSA} in the following form
\begin{equation}\label{averaged T after SSA}
  \avg{\con{T}}=\sum_n \avg{\con{Q}_n}=\sum_n\avg{\con{t}_n} + \sum_{n\neq m} \avg{\con{t}_n} \bar{\con{G}} \avg{\con{t}_m} + \cdots.
\end{equation}
As an immediate result, the CPA self-consistent condition $\avg{T}=0$ is simplified to
\begin{equation}\label{t n equals zero}
  \avg{\con{t}_n} \equiv \sum_Q c_n^Q \con{t}_n^Q = 0,
\end{equation}
where $c^Q_n$ is the concentration of Q element on the site $n$. Combining the above single-site equation and Eq.\eqref{Dyson equation for effective medium Green's function}, the on-site self-energy $\con{\Sigma}_{n,im}$ can be self-consistently solved for each site of the system.
In such a way, the effective medium described by $\con{\Sigma}_{im} =\sum_n \con{\Sigma}_{n,im}$ can be efficiently obtained.
By here, we have introduced the central idea of generalized CPA with the single site approximation.

\subsection{Application to Keldysh's Representation}
The quantities we defined so far in the generalized CPA with SSA (such as $\con{G}$, $\con{\Sigma}$, $\con{V}$, $\con{T}$ and their single-site counterparts) are all defined for a general case. If we substitute with the retarded/advanced quantities, we obtain the conventional CPA formalism. Here, we apply the generalized CPA to the Keldysh's real-time matrix representation, aiming to obtain the disorder average of all the real-time single-particle GFs introduced in Sec.\ref{sec:A Brief Review on the Green's Function Method}.
To do this, we need to rewrite the quantities $\con{G}$, $\con{\Sigma}$, $\con{V}$, $\con{T}$ and their single-site counterparts in the form of the Keldysh's matrix in Eq.\eqref{Keldysh 2-by-2 matrix}. For example,
$
\con{\Sigma} =
\begin{tiny}
  \left(
    \begin{array}{cc}
      \Sigma^A & 0 \\
      \Sigma^K & \Sigma^R\\
    \end{array}
  \right)
\end{tiny},
$
$
\con{T} =
\begin{tiny}
  \left(
    \begin{array}{cc}
      T^A & 0 \\
      T^K & T^R\\
    \end{array}
  \right)
\end{tiny},
$
and $
\con{V} =
\begin{tiny}
  \left(
    \begin{array}{cc}
      V^A & 0 \\
      V^K & V^R \\
    \end{array}
  \right)
\end{tiny}
=\begin{scriptsize}
  \left(
    \begin{array}{cc}
      V & 0 \\
      0 & V \\
    \end{array}
  \right)
\end{scriptsize}
$ since the potential is Hermitian and can take its simpler form. Replacing the quantities in Eq.\eqref{Dyson equation for effective medium Green's function} with Keldysh's matrices leads to the following equations (see more details in Append. \ref{app:GCPA Calculation Decoupling}),
\begin{subequations}\label{GCPA first step decoupled}
\begin{align}
  \bar{G}^R &= G_0^R({I} - \Sigma^R G_0^R)^{-1},\label{GCPA first step decoupled 1}\\
  \bar{G}^A &= G_0^A({I} - \Sigma^A G_0^A)^{-1},\label{GCPA first step decoupled 2}\\
  \bar{G}^K &= \bar{G}^R \Sigma^K \bar{G}^A + (I + \bar{G}^R\Sigma^R)G_0^K (I + \Sigma^A \bar{G}^A)\label{GCPA first step decoupled 3}.
\end{align}
\end{subequations}
where $\Sigma=\Sigma_{ld} + \Sigma_{im}$.
Eqs.(\ref{GCPA first step decoupled 1},\ref{GCPA first step decoupled 2}) for retarded and advanced GFs are the same as the conventional CPA.
Eq.\eqref{GCPA first step decoupled 3} is usually called the Keldysh's formula for $G^K$ which relates $\bar{G}^K$ to $\bar{G}^{R/A}$ and different components of $\Sigma$.
From the above equation, we can see the three components of $G$, namely $G^{R/A/K}$, are not independent of each other: $G^A$ is the conjugate of ${G^R}$, and thus $G^K$ is given by $G^R$ through the Keldysh's formula. Thus $G^R$ provides the sufficient knowledge to compute NEGFs, provided the self-energy $\Sigma$.  This fact forms the important physical foundation for the CPA-NVC method in which conventional CPA is carried out only for $\bar{G}^{R/A}$. Actually, the similar relations between the retarded/advanced and the Keldysh quantities can also be found for other quantities, such as $T$ and $t_n$ as we show in the following.

To obtain the CPA equations, we apply the Keldysh's matrices to Eq.\eqref{t-matrix expansion}, and find
\begin{subequations}\label{GCPA second step decoupled for T}
\begin{align}
  T^R =& [I - (V- \Sigma_{im}^R)\bar{G}^R]^{-1}(V - \Sigma_{im}^R),\label{GCPA second step decoupled 1 for T}\\
  T^A =& [I - (V - \Sigma_{im}^A)\bar{G}^A]^{-1}(V - \Sigma_{im}^A),\label{GCPA second step decoupled 2 for T}\\
  T^K =& T^R \bar{G}^K T^A - (I + T^R \bar{G}^R) \Sigma_{im}^K (I + \bar{G}^A T^A).\label{GCPA second step decoupled 3 for T}
\end{align}
\end{subequations}
Similarly, applying the Keldysh's matrices to Eq.\eqref{definetn} leads to
\begin{subequations}\label{GCPA second step decoupled}
\begin{align}
  t_n^R =& [I - (v_n - \Sigma_{n,im}^R)\bar{G}^R]^{-1}(v_n - \Sigma_{n,im}^R),\label{GCPA second step decoupled 1}\\
  t_n^A =& [I - (v_n - \Sigma_{n,im}^A)\bar{G}^A]^{-1}(v_n - \Sigma_{n,im}^A),\label{GCPA second step decoupled 2}\\
  t_n^K =& t_n^R \bar{G}^K t_n^A - (I + t_n^R \bar{G}^R) \Sigma_{n,im}^K (I + \bar{G}^A t_n^A).\label{GCPA second step decoupled 3}
\end{align}
\end{subequations}
The quantity $\Sigma^K_{im}$ is called the non-equilibrium coherent potential in the literatures of NECPA \cite{PhysRevB.85.235111,PhysRevB.88.205415}. Here the Eqs.(\ref{GCPA second step decoupled 3 for T} ,\ref{GCPA second step decoupled 3}) can be called the Keldysh's formula for $T$ and $t_n$.
Similar to $G$, we find that the retarded, advanced and Keldysh components of $T$ or $t_n$ are also not independent. Given a self-energy, the retarded quantity can determine the other two components, providing a foundation for the equivalence of CPA-NVC and NECPAs as we will see in next section.
After applying to the Keldysh's representation, by combining Eqs.\eqref{GCPA first step decoupled} and Eqs.\eqref{GCPA second step decoupled} with the CPA condition $\avg{t_n^{R/A/K}}=0$ in SSA, we can self-consistently compute the self-energy $\Sigma_{im}^{R/A/K}$ that gives $\bar{G}^{R/A/K}$ of the effective medium. As an important result, according to the relations in Append. \ref{app:Expressing various real-time Green's functions}, the average of all other real-time single-particle GFs can be easily obtained.

\section{Generalized Non-equilibrium Vertex Correction}\label{sec:Generalized Vertex Correction}
The generalized CPA only provides a way to average a single-particle GF.
However, many physical quantities contain the product of two GFs, such as the quantum transport properties mentioned in Sec.\ref{sec:Quantum transport properties of disordered device}.
Because the two GFs describing the same disordered system are internally correlated, $\avg{GCG}$ is not simply equal to $\avg{G}C\avg{G}$ where C is an arbitrary constant.
For this reason, a new algorithm called the generalized NVC is formulated in this section to correctly compute $\avg{GCG}$, so that the disorder averaged product of any two real-time GFs can be obtained, such as $\avg{\con{G}^< C \con{G}^<}$.

\subsection{Theory of generalized non-equilibrium vertex correction}
Here, we consider a two-GF correlator
\begin{equation}\label{GVC problem}
  K = \avg{\con{G}(z_1) C \con{G}(z_2)},
\end{equation}
where $C$ is an arbitrary constant.
In Eq.\eqref{GVC problem}, the GFs can be at two different energies.
For simplicity, these energy indices will be suppressed in the rest of the derivation.
To evaluate $K$, we insert Eq.\eqref{T matrix equation} into Eq.\eqref{GVC problem} and apply the CPA condition $\avg{T}=0$, and then obtain
\begin{equation}\label{vertex correction}
  \avg{\con{G} C \con{G}} = \bar{\con{G}}(C + \Omega) \bar{\con{G}},
\end{equation}
where
\begin{equation}\label{vertex correction2}
 \Omega \equiv \avg{\con{T} \bar{\con{G}} C \bar{\con{G}} \con{T}}
\end{equation}
is the generalized NVC, containing all the effects of disorders on the two-GF correlator.

\begin{figure}
  \centering
  % Requires \usepackage{graphicx}
  \includegraphics[width=8.4cm]{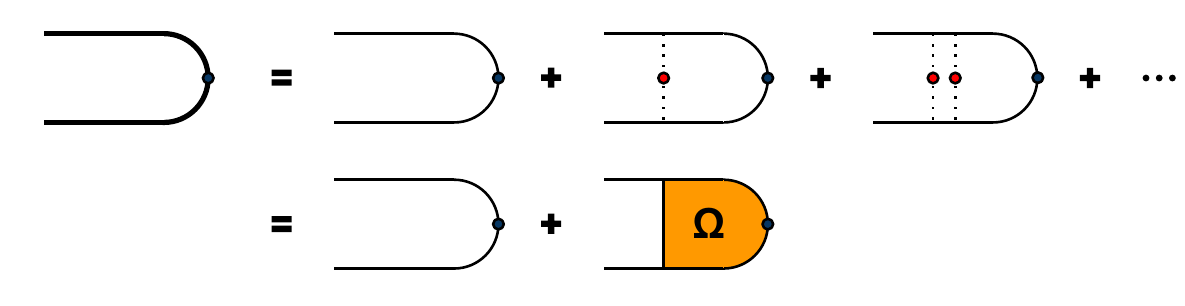}\\
  \caption{(color online) Diagrammatic representation of $\avg{\con{G} C \con{G}}$. The thick line and thin line represent $\con{G}$ and $\bar{\con{G}}$ respectively. The dash line represents the interaction with the disorders (red dots). The blue dot represents a vertex $C$.}\label{fig: feynamn diagram}
\end{figure}

In order to compute $\Omega$, we substitute the $T$ with Eq.\eqref{T in terms of Q} and then obtain
\begin{equation}\label{Gamma in terms of Q}
  \Omega = \sum_n \sum_m\avg{\con{Q}_n \bar{\con{G}} C \bar{\con{G}} \tilde{\con{Q}}_m}.
\end{equation}
For terms with $n\neq m$, by applying SSA, we can obtain $\avg{\con{Q}_n \bar{\con{G}} C \bar{\con{G}} \tilde{\con{Q}}_m} = 0$.
Consequently, Eq.\eqref{Gamma in terms of Q} is simplified to
\begin{equation}\label{Gamma as the sum of Gamma n}
  \Omega = \sum_n \Omega_n,
\end{equation}
where we have defined $\Omega_n \equiv \avg{\con{Q}_n \bar{\con{G}} C \bar{\con{G}} \tilde{\con{Q}}_n}$.
To proceed further, we replace the $\con{Q}_n$, $\tilde{\con{Q}}_n$ with the relation in Eq.\eqref{Q iterative}, $\con{Q}_n = \con{t}_n (\con{I} + \bar{\con{G}} \sum_{p\neq n} \con{Q}_p)$ and its counterpart $\tilde{\con{Q}}_n = (\con{I} + \sum_{q\neq n} \tilde{\con{Q}}_q\bar{\con{G}})\con{t}_n$, and get
\begin{equation}\label{Gamma n replaced}
  \Omega_n = \avg{\con{t}_n(\con{I}+\bar{\con{G}} \sum_{p\neq n}\con{Q}_p)\bar{\con{G}} C \bar{\con{G}}(\con{I}+\sum_{q\neq n}\tilde{\con{Q}}_q \bar{\con{G}} ) \con{t}_n}.
\end{equation}
Expanding the products in $\avg{\cdots}$, we will get four terms,
among which, after applying SSA, two terms involving only one $\con{Q}$ vanish , and the term involving the product of two $\con{Q}$s is simplified to $\avg{\con{t}_n \bar{\con{G}} \sum_{p\neq n} \Omega_p \bar{\con{G}} \con{t}_n}$.
Therefore, Eq.\eqref{Gamma n replaced} finally becomes
\begin{equation}\label{linear eq for gamma n}
  \Omega_n = \avg{\con{t}_n \bar{\con{G}} C \bar{\con{G}} \con{t}_n} + \sum_{p\neq n}\avg{\con{t}_n \bar{\con{G}} \Omega_p\bar{\con{G}} \con{t}_n},
\end{equation}
which forms a closed set of linear equations for the unknown $\Omega_n$. In Eq.\eqref{linear eq for gamma n}, the average is over pairs
of scattering events on the same site. In other words the scattering from different sites is regarded as statistically
uncorrelated and the motion of two particles, represented by the two GFs, in the medium is correlated only if they both scatter from the same site.
Solving Eq.\eqref{linear eq for gamma n} leads to $\Omega_n$ for each disordered site, and thus the averaged two-GF correlator in Eq.\eqref{GVC problem} can be obtained.
The procedure to average the two-GF correlator from Eq.\eqref{vertex correction} to Eq.\eqref{linear eq for gamma n} can be represented by the Feynman diagrams as shown in Fig.\ref{fig: feynamn diagram}.\cite{PhysRevB.2.1771} The first line in Fig.\ref{fig: feynamn diagram} expresses the two-GF correlator with an infinite series of ladder diagrams that refers to the direct expansion of the GFs in SSA, and the second line reduces the infinite ladder series to a single NVC.
With this simple Feynman diagram, the various two-GF correlators, such as $G^< C G^<$, can be calculated in a much more efficient way than the method reported in Ref.\onlinecite{PhysRevB.88.085420}.

\subsection{Application to Keldysh's Representation}
Similar to the generalized CPA, the generalized NVC formalism can also be applied to Keldysh's representation.
To do so, we first consider the arbitrary constant $C$ matrix for the following four different cases
\begin{eqnarray*}
  C^{(1)} = \left(
                \begin{array}{cc}
                  C & 0 \\
                  0 & 0 \\
                \end{array}
              \right),
  \qquad
  C^{(2)} = \left(
                \begin{array}{cc}
                  0 & C \\
                  0 & 0 \\
                \end{array}
              \right),\\
  C^{(3)} = \left(
                \begin{array}{cc}
                  0 & 0 \\
                  C & 0 \\
                \end{array}
              \right),
  \qquad
  C^{(4)} = \left(
                \begin{array}{cc}
                  0 & 0 \\
                  0 & C \\
                \end{array}
              \right).
\end{eqnarray*}
By applying these four $C^{(i)}$s to Eq.\eqref{vertex correction2}, we obtain four different $\Omega^{(i)}$s as follows,
\begin{eqnarray*} \label{ninegammas}
  {\Omega}^{(1)} &= \left(
                     \begin{array}{cc}
                       \Omega^{AA} & 0 \\
                       \Omega^{KA} & 0 \\
                     \end{array}
                   \right),
  \qquad
  {\Omega}^{(2)} &= \left(
                     \begin{array}{cc}
                       \Omega^{AK} & \Omega^{AR} \\
                       \Omega^{KK} & \Omega^{KR} \\
                     \end{array}
                   \right),\\
  {\Omega}^{(3)} &= \left(
                     \begin{array}{cc}
                       0 & 0 \\
                       \Omega^{RA} & 0 \\
                     \end{array}
                   \right),
  \qquad
  {\Omega}^{(4)} &= \left(
                     \begin{array}{cc}
                       0 & 0 \\
                       \Omega^{RK} & \Omega^{RR} \\
                     \end{array}
                   \right).
\end{eqnarray*}
Applying these $\Omega^{(i)}$ and the corresponding $C^{(i)}$ to Eq.\eqref{vertex correction} leads to nine different pairwise combinations of $G^R$, $G^A$ and $G^K$
\begin{eqnarray}
  \avg{G^R C G^R} &=& \bar{G}^R (C + \Omega^{RR}) \bar{G}^R,\\
  \avg{G^R C G^A} &=& \bar{G}^R (C + \Omega^{RA}) \bar{G}^A,\\
  \avg{G^A C G^R} &=& \bar{G}^A (C + \Omega^{AR}) \bar{G}^R,\\
  \avg{G^A C G^A} &=& \bar{G}^A (C + \Omega^{AA}) \bar{G}^A,\\
  \avg{G^R C G^K} &=& \bar{G}^R \Omega^{RK} \bar{G}^A + \bar{G}^R (C + \Omega^{RR}) \bar{G}^K,\\
  \avg{G^A C G^K} &=& \bar{G}^A \Omega^{AK} \bar{G}^A + \bar{G}^A (C + \Omega^{AR}) \bar{G}^K,\\
  \avg{G^K C G^R} &=& \bar{G}^R \Omega^{KR} \bar{G}^R + \bar{G}^K (C + \Omega^{AR}) \bar{G}^R,\\
  \avg{G^K C G^A} &=& \bar{G}^R \Omega^{KA} \bar{G}^A + \bar{G}^K (C + \Omega^{AA}) \bar{G}^A,\\
  \avg{G^K C G^K} &=& \bar{G}^R \Omega^{KK} \bar{G}^A + \bar{G}^K \Omega^{AK} \bar{G}^A\nonumber\\
    && + \bar{G}^R \Omega^{KR}\bar{G}^K + \bar{G}^K (C + \Omega^{AR})\bar{G}^K.
\end{eqnarray}
The linear combination of these nine quantities can give all the real-time two-GF correlators (see Append.\ref{app:Expressing various real-time Green's functions} for more details).
For example:
\begin{equation}
\begin{split}
  \avg{G^< C G^<} = \frac{1}{4} \Big[  &\avg{G^R C G^R} - \avg{G^R C G^A} - \avg{G^R C G^K}\\
                                     - &\avg{G^A C G^R} + \avg{G^A C G^A} + \avg{G^A C G^K}\\
                                     - &\avg{G^K C G^R} + \avg{G^K C G^A} + \avg{G^K C G^K}\Big].
\end{split}
\end{equation}

The remaining task is to find the nine generalized NVC quantities defined in the four $\Omega^{(i)}$s.
By inserting the Keldysh's matrices into Eq.\eqref{linear eq for gamma n}, we obtain nine linear equations with details provided in the Append.\ref{app:Nine Equations for GNVCs}.
From Append.\ref{app:Nine Equations for GNVCs}, we can see some of these quantities are coupled with each other.
However, Solving these linear equations from top to down leads to the decoupling of the calculation, giving a unified solution for these 9 generalized NVCs.
Therefore, with the generalized NVC, the averaged physical properties which contain two Green's function correlators, such as averaged non-equilibrium electron density, averaged current, current fluctuation and averaged shot noise (see Sec.\ref{sec:Quantum transport properties of disordered device}), can all be computed in a unified and efficient way.

\subsection{Relation between NECPA and CPA-NVC}
In this subsection, we will clarify the internal relation between NECPA\cite{PhysRevB.85.235111, PhysRevB.88.205415} and CPA-NVC\cite{PhysRevLett.100.166805} by taking a close look at the non-equilibrium coherent potential defined in NECPA, which is the quantity $\Sigma^K_{im}$ in this paper.
We start from the second term of the Keldysh's formula in Eq.\eqref{GCPA first step decoupled 3}.
By using Dyson equation Eq.\eqref{Dyson equation} for $G^{R/A}$ and fluctuation-dissipation theorem\cite{haug2008quantum}, we get
\begin{equation}\label{bound states}
\begin{split}
  (I + \bar{G}^R& \Sigma^R)G_0^K (I + \Sigma^A \bar{G}^A)\\
       &= [1-2f(E)] G^R [(G_0^A)^{-1} - (G_0^R)^{-1}]G^A.
\end{split}
\end{equation}
Since $(G_0^A)^{-1} - (G_0^R)^{-1} = -2 i \eta$ where $\eta \rightarrow 0$, thus Eq.\eqref{bound states} will equal zero except $G^{R/A}(E)$ diverges, which means the energy $E$ coincidentally equals the bound-state energy.\cite{PhysRevB.79.045119}
Furthermore, this term when it is nonzero is only relevant to the initial transient of time dependent problem, and thus it can be neglected for the steady-state problem that we are working on here (see discussions in page 305 in Ref.\onlinecite{datta1997electronic} and Eq.2.16 in Ref.\onlinecite{PhysRevB.36.2578}).
The great success of NEGF-DFT based quantum transport methods \cite{PhysRevB.63.121104,PhysRevB.63.245407,PhysRevB.71.195422,PhysRevLett.96.166804,:/content/aip/journal/jcp/115/9/10.1063/1.1391253,Xue2002151,PhysRevB.65.165401,PhysRevB.70.085410,Rocha2005,PhysRevLett.95.206805} further confirms this fact with great amount of practical applications.
After dropping the bound-state term, the Keldysh's Green's function $\bar{G}^K$ becomes
\begin{equation}\label{Keldysh for K}
  \bar{G}^K = \bar{G}^R \Sigma^K \bar{G}^A.
\end{equation}

Based on Eq.\eqref{Keldysh for K}, we can prove the equivalence of the NECPA and CPA-NVC methods.
At first, we consider Eq.\eqref{GCPA second step decoupled 3 for T} for $T^K$. By applying the generalized CPA self-consistent condition $\avg{T}=0$, namely $\avg{T^{R/A/K}}=0$, we immediately obtain the nonequilibrium coherent potential in the following form
\begin{equation}
  \Sigma_{im}^K = \avg{T^R \bar{G}^R \Sigma_{ld}^K \bar{G}^A T^A},
\end{equation}
which is exactly the same as the NVC in Eq.\eqref{vertex correction2}.
Applying the SSA still don't change the conclusion. In particular, applying $\avg{t_{n}^{K}}=0$ to Eq.\eqref{GCPA second step decoupled 3} results in
\begin{equation*}
  \Sigma_{n,im}^K = \avg{t_n^R \bar{G}^R \Sigma_{ld} \bar{G}^A t_n^A} + \sum_{m \neq n} \avg{t_n^R \bar{G}^R \Sigma_{m,im}^K \bar{G}^A t_n^A},
\end{equation*}
which is the same as the NVC in Eq.\eqref{linear eq for gamma n} after SSA.
By here, we have shown the non-equilibrium coherent potential equals NVC and provides no more physics.
For these reasons, the NECPAs introduced in the Refs.\onlinecite{PhysRevB.85.235111,PhysRevB.88.205415} are essentially the same as the CPA-NVC. This equivalence is guaranteed by the Keldysh's formulas in Eqs.\eqref{GCPA first step decoupled 3},\eqref{GCPA second step decoupled 3 for T} and \eqref{GCPA second step decoupled 3}
which are inherently incorporated in the NECPA, but are explicitly used in the derivation of CPA-NVC.

\section{Realizing First-Principles Calculation}\label{sec:First Principle Calculation}
In previous sections, we have introduced the generalized CPA-NVC algorithm to treat the disorder effects in non-equilibrium quantum transport. However, the solution of the generalized CPA and NVC equations need us to provide the potential $v_n^Q$ of each $Q$ element on the site $n$, namely the electronic structure of the disordered device. In this section, we will discuss how to combine the generalized CPA-NVC with NEGF-DFT method to calculate the non-equilibrium electronic structure of the disordered nanoelectronics from first principles.

\subsection{Conditionally averaged Green's function}

\begin{figure}
  \centering
  % Requires \usepackage{graphicx}
  \includegraphics[width=7.0cm]{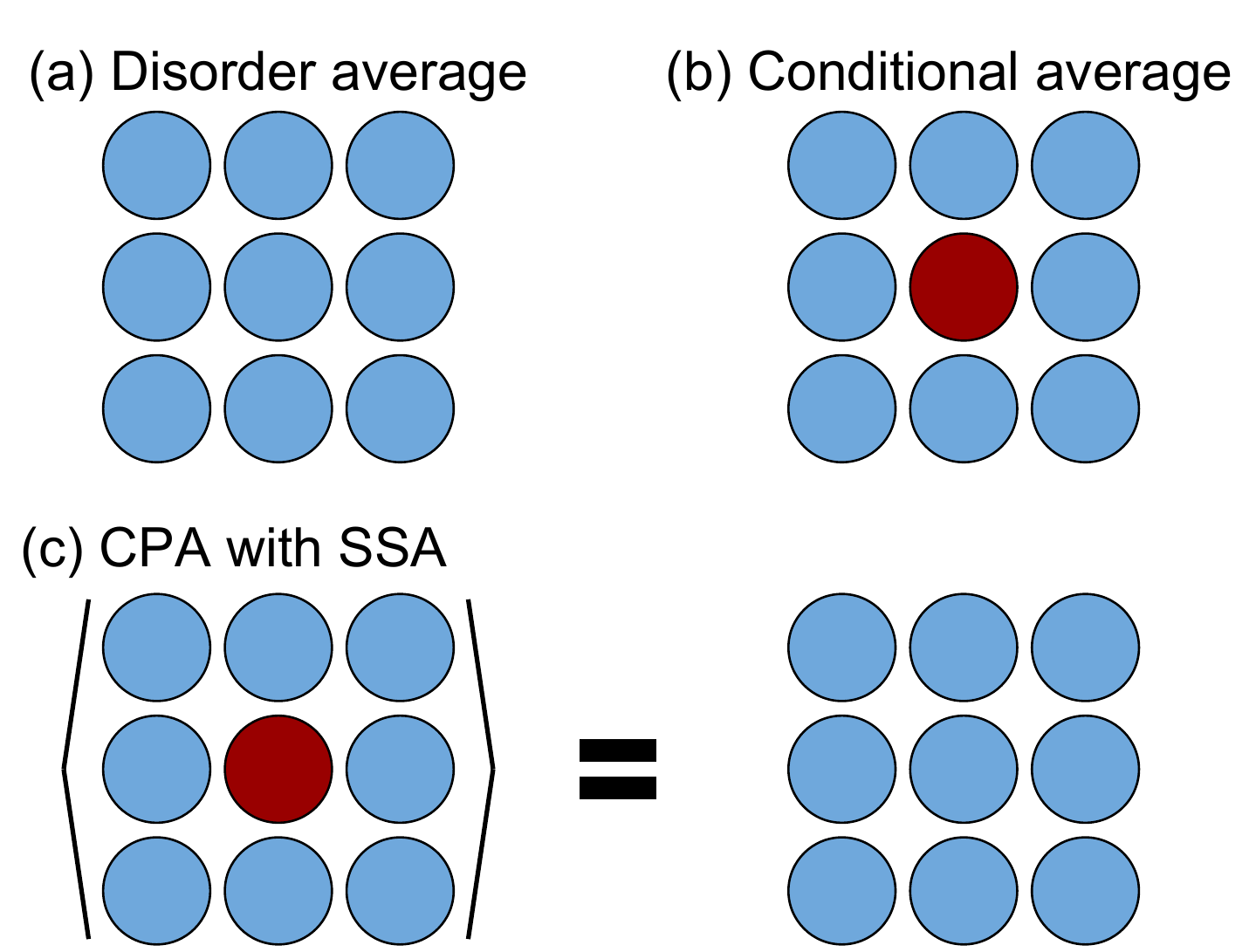}\\
  \caption{(color online) (a) A fully disorder averaged system. (b) A conditionally averaged system.
   (c) Schematic illustration of CPA with SSA.}\label{fig: conditional}
\end{figure}

The central quantity for realizing DFT self-consistent calculation is the conditionally averaged lesser Green's function $\bar{G}^{<,Q}$, which gives the $\rho_n^Q$ to update $v_n^Q$ in each DFT iteration.
In general, the conditionally averaged GF $\bar{G}^{Q}$ is associated with the system in which the n-th site is occupied by the fixed $Q$ element, and the disorder average is carried out for the rest of the disordered sites.
Thus, $\bar{G}^{Q}$ corresponds to the effective medium with $Q$ element embedded on the site n, as shown in  Fig.\ref{fig: conditional}(b).
In order to calculate $\con{\bar{G}}^Q$, we expand it with reference to $\con{\bar{G}}$ shown in Fig.\ref{fig: conditional}(a) by using Eq.\eqref{T matrix equation}, and obtain
\begin{equation}\label{conditional averaged G}
  \bar{\con{G}}^Q = \bar{\con{G}} + \bar{\con{G}} \con{t}_n^{Q} \bar{\con{G}},
\end{equation}
where
\begin{equation}\label{on site t n Q}
  \con{t}_n^Q = [\con{I} - (\con{v}^Q_n - \con{\Sigma}_{n,im})\bar{\con{G}}]^{-1}(\con{v}_n^Q - \con{\Sigma}_{n,im}).
\end{equation}
Note that we have used $T=t_n^Q$ since there is only one scattering center. One can check that
\begin{equation}\label{alternative cpa}
  \sum_Q c^Q \bar{G}^Q = \bar{G}
\end{equation}
by applying the single-site CPA condition $\avg{t_n} = 0$ in Eq.\eqref{conditional averaged G}.
Fig.\ref{fig: conditional}(c) provides a schematic illustration of the above equation.
By substituting with Keldysh's matrices in Eq.\eqref{conditional averaged G}, we obtain
\begin{subequations}\label{conditional averaged G for calculation}
\begin{eqnarray}
  \bar{\con{G}}^{R,Q} &=& \bar{G}^R + \bar{G}^R t_n^{R,Q} \bar{G}^R, \\
  \bar{\con{G}}^{A,Q} &=& \bar{G}^A + \bar{G}^A t_n^{A,Q} \bar{G}^A, \\
  \bar{\con{G}}^{K,Q} &=& \bar{G}^K + \bar{G}^R t_n^{K,Q} \bar{G}^A \nonumber \\
                        && \qquad+ \bar{G}^K t_n^{A,Q} \bar{G}^A + \bar{G}^R t_n^{R,Q} \bar{G}^K.
\end{eqnarray}
\end{subequations}
where the matrices $t^{R/A/K}_n$ are defined in Eq.\eqref{GCPA second step decoupled}.
With above three conditionally averaged GFs, $\bar{G}^{<,Q}$ can be calculated by the relation
\begin{equation}\label{Gless conditionally averaged}
  \bar{G}^{<,Q} = \frac{1}{2}(-\bar{G}^{R,Q} + \bar{G}^{A,Q} + \bar{G}^{K,Q}).
\end{equation}
The conditionally averaged $\bar{G}^{<,Q}$ provides the non-equilibrium density matrix $\bar{\rho}^Q$ for each disordered element in the system. In combination with DFT, the potential $v_n^Q$ can be computed from the electron density. Consequently, the non-equilibrium electronic structure of the disordered nanoelectronics can be self-consistently calculated by combining NEGF-DFT with generalized CPA-NVC method. As a result, the effects of disorders on the quantum transport properties can be simulated from atomistic first principles.

\subsection{Self-consistent procedures}
\begin{figure}
  \centering
  \includegraphics[width=8.4cm]{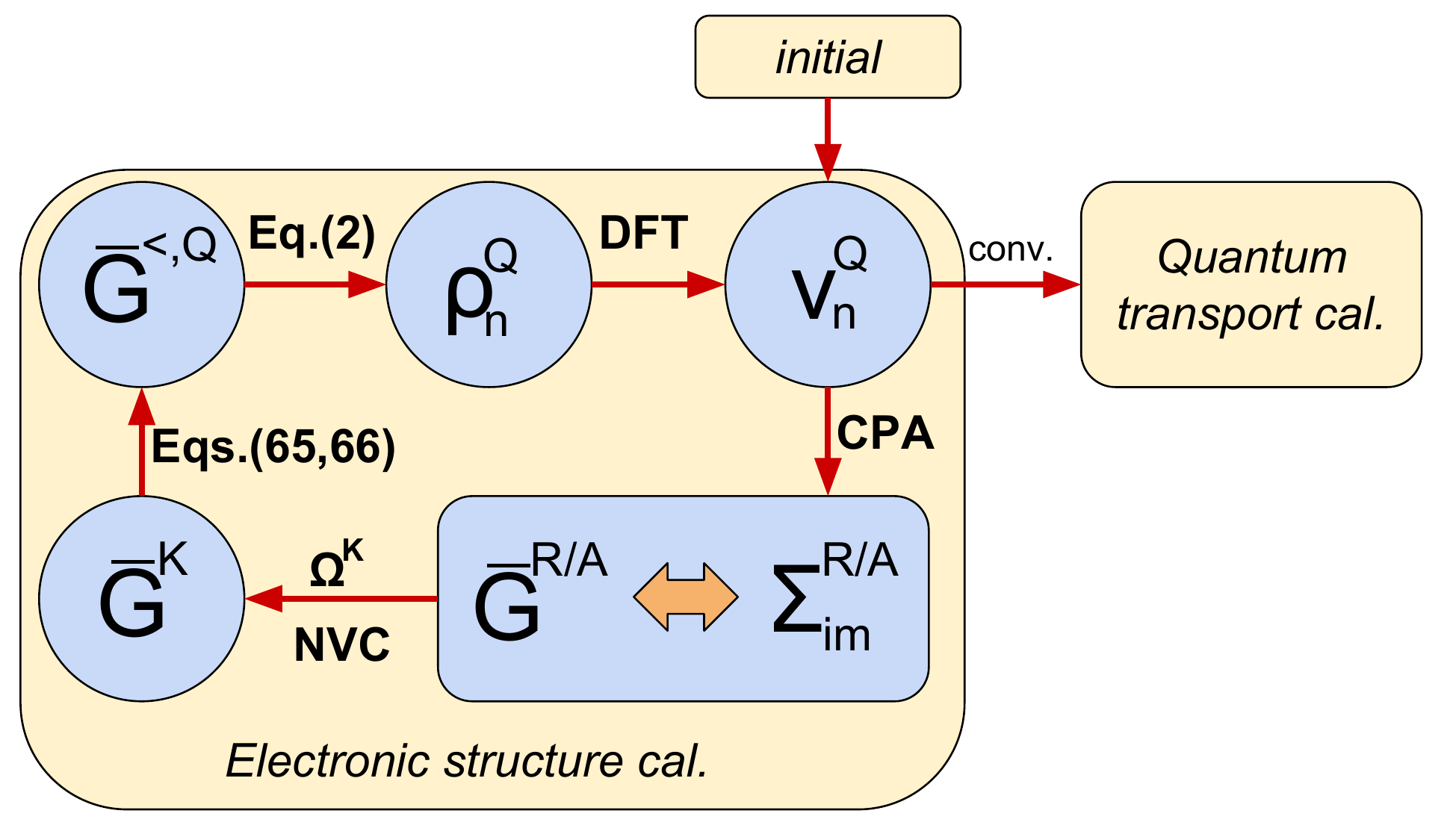}\\
  \caption{(color online) Flowchart for implementing the generalized CPA-NVC within the NEGF-DFT method to realize first principles simulation of disordered nano-electronics.}\label{fig: GCPA-loop}
\end{figure}

Here, we briefly summarize the major procedures for implementing the generalized CPA-NVC within the framework of NEGF-DFT. As shown Fig.\ref{fig: GCPA-loop}, the whole self-consistent calculation involves the following important steps:
(i) For the given device geometry, compositions and their concentrations on each site, and the left and right leads, we choose an appropriate initial atomic potential $ v_n^{Q}$ for each atomic species. This potential can be constructed by a self-consistent calculation of a single atom or a bulk phase of the element. In addition, one has to calculate the self-energy of the left and right leads, e.g., $\Sigma_{ld}$, which are then kept constant during the self-consistent calculation.
(ii) With $v_n^Q$, we solve conventional CPA nonlinear equations self-consistently to obtain the coherent potential $\Sigma_{im}^R$ that directly leads to the averaged retarded GF $\bar{G}^{R}$.
(iii) By using the matrix elements of $\bar{G}^{R}$ associated with the disordered sites, we can solve the NVC linear equation Eq.\eqref{linear eq for gamma n} to obtain the $\Omega_{\mathrm{NVC}}$ that gives $\bar{G}^K$.
(iv) With $\bar{G}^{R/A/K}$, we calculate the configurationally averaged $\bar{G}^{R/A/K,Q}$ in Eqs.\eqref{conditional averaged G for calculation} to obtain $\bar{G}^{<,Q}$ with Eq.\eqref{Gless conditionally averaged}.
(v) We obtain the electron density $\rho_n^Q$ for each element of the device from their conditionally averaged non-equilibrium density matrix given by $\bar{G}^{<,Q}$.
(vi) We update the electronic potential $v^{Q}_n$ with DFT and check if the potential $v_n^Q$ is converged for each element in the system. If not, we start a new iteration by going back to step (ii) with the updated $v^{Q}_{n}$.
Such an iterative procedure continues until the electronic potential is converged.
With the converged non-equilibrium electronic structure, we calculate the averaged quantum transport properties with the generalized CPA-NVC, such as an $I-V$ curve, current fluctuation and averaged shot noise, to finish the simulation of a disordered nano-electronics.

\section{Conclusion} \label{sec: Conclusion}
We have developed a generalized CPA-NVC formalism to realize quantum transport simulation of disordered nanoelectronic devices at non-equilibrium state. Based on the generalized CPA-NVC formalism, We show that the averaged product of any two real-time single-particle GFs can be computed with nine generalized NVCs, which account for the multiple impurity scattering and non-equilibrium quantum statistics. As an important result, various non-equilibrium quantum transport properties, including averaged non-equilibrium density matrix, averaged current, current fluctuation and averaged shot noise, can all be effectively computed with a unified scheme. Moreover, We clarify that the equivalence between NECPA and CPA-NVC is guaranteed by the Keldysh's formulas and the non-equilibrium coherent potential equals NVC. In addition, the generalized conditionally averaged NEGF is derived to combine with DFT to enable first principles simulation of disordered nanoelectronics.
As a summary, Our approach provides a unified, efficient and self-consistent method for simulating non-equilibrium quantum transport through disordered nanoelectronics.

\begin{acknowledgments}
 This work is supported by ShanghaiTech University start-up fund (Y. Ke). We thank Prof. Hong Guo at McGill Univeristy for reading our manuscript and providing important comments on our work.
\end{acknowledgments}

\appendix

\section{Relations between Analytical Continuation with the Langreth Theorem and the Matrix Representation}\label{app:Relations between the Langreth Theorem and the Matrix Representation}
The operation on contour-ordered quantities involves integrating along the time-contour, which can be transformed to integrating along the real-time axis by using the Langreth theorem.\cite{haug2008quantum}
For example, suppose $\con{A}$, $\con{B}$ and $\con{D}$ are three quantities defined on the contour and have the relation
\begin{equation}\label{D equals AB}
  \con{D} = \con{A} \con{B}.
\end{equation}
According to the Langreth theorem, their real-time counterparts have the relations
\begin{eqnarray}
  D^< &=& A^R B^< + A^< B^A,\label{Langreth 1}\\
  D^R &=& A^R B^R.\label{Langreth 2}
\end{eqnarray}
These two identities can be derived by deforming the time contour as indicated in Fig.4.4 in Ref.\onlinecite{haug2008quantum}. An alternative way to apply the Langreth theorem is by using the Craig's or Keldysh's 2-by-2 real-time matrix representation of the contour-ordered quantities. For example, by substituting the matrix notation defined in \eqref{Craig 2-by-2 matrix} into \eqref{D equals AB}, we obtain
\begin{equation*}
\begin{split}
  \left(
          \begin{array}{cc}
            D^t & -D^< \\
            D^> & -D^{\bar{t}} \\
          \end{array}
        \right)
        =
        \left(
          \begin{array}{cc}
            A^t B^t - A^< B^> & -A^tB^< + A^< B^{\bar{t}} \\
            A^> B^t - A^{\bar{t}}B^> & -A^>B^< + A^{\bar{t}}B^{\bar{t}} \\
          \end{array}
        \right).
\end{split}
\end{equation*}
From this expression, we directly recover \eqref{Langreth 1}
\begin{equation*}
\begin{split}
  D^< &= A^tB^< - A^< B^{\bar{t}}\\
   &= (A^t - A^<)B^< + A^<(B^< - B^{\bar{t}}) = A^RB^< + A^<A^A.
\end{split}
\end{equation*}
Similarly, by substituting Eq.\eqref{Keldysh 2-by-2 matrix} into Eq.\eqref{D equals AB} leads to Eq.\eqref{Langreth 2}.
Therefore, we can regard these 2-by-2 matrices inherently incorporate the Langreth theorem and are preferred using in practice.

\section{Derivation of Eq.\eqref{GCPA first step decoupled}}\label{app:GCPA Calculation Decoupling}
We firstly rewritten Eq.\eqref{Dyson equation for effective medium Green's function} as the explicit form for $\con{\bar{G}}$ that
\begin{equation}
  \con{\bar{G}} = \con{G}_0 (\con{I} - \con{\Sigma}\con{G}_0)^{-1}.
\end{equation}
By replacing with the Keldysh's representation defined in Eq.\eqref{Keldysh 2-by-2 matrix}, we obtain
\begin{equation}
  \left(
    \begin{array}{cc}
      \bar{G}^A & 0 \\
      \bar{G}^K & \bar{G}^R \\
    \end{array}
  \right)
  =
  \left(
    \begin{array}{cc}
      G_0^A & 0 \\
      G_0^K & G_0^R \\
    \end{array}
  \right)
  \left(
    \begin{array}{cc}
      A & 0 \\
      K & R \\
    \end{array}
  \right)^{-1},
\end{equation}
where we have defined
\begin{eqnarray}
  R &\equiv& I - \Sigma^R G_0^R,\\
  A &\equiv& I - \Sigma^A G_0^A,\\
  K &\equiv& -\Sigma^K G_0^A - \Sigma^R G_0^K.
\end{eqnarray}
Using the identity
\begin{equation}\label{inversion 2by2}
  \left(
    \begin{array}{cc}
      A & 0 \\
      K & R \\
    \end{array}
  \right)^{-1}
  =
  \left(
    \begin{array}{cc}
      A^{-1} & 0 \\
      -R^{-1}KA^{-1} & R^{-1} \\
    \end{array}
  \right),
\end{equation}
then we can get
\begin{subequations}
\begin{align}
  \bar{G}^R &= G_0^R R^{-1} = G_0^R[{I} - \Sigma^R G_0^R]^{-1},\\
  \bar{G}^A &= G_0^A A^{-1} = G_0^A[{I} - \Sigma^A G_0^A]^{-1},\\
  \bar{G}^K &= G_0^K A^{-1} - G_0^R R^{-1} K A^{-1}\nonumber\\
            &= \bar{G}^R \Sigma^K \bar{G}^A + (I + \bar{G}^R \Sigma^R)G_0^K (G_0^A)^{-1} \bar{G}^A\nonumber\\
            &= \bar{G}^R \Sigma^K \bar{G}^A + (I + \bar{G}^R\Sigma^R)G_0^K (I + \Sigma^A \bar{G}^A).
\end{align}
\end{subequations}

\section{Expressing various real-time quantities in terms of $Q^R$,$Q^A$ and $Q^K$}\label{app:Expressing various real-time Green's functions}
This appendix provides a convenient way to express the various real-time quantities in terms of the linear combinations of $Q^R$, $Q^A$ and $Q^K$ by using the Keldysh's linear transformation shown as follows:
\begin{equation*}
\begin{split}
  \left(
    \begin{array}{cc}
      Q^t & -Q^< \\
      Q^> & -Q^{\bar{t}} \\
    \end{array}
  \right)
  =\frac{1}{2}
  \left(
    \begin{array}{cc}
      1 & 1 \\
      -1 & 1 \\
    \end{array}
  \right)
  \left(
    \begin{array}{cc}
      Q^A & 0 \\
      Q^K & Q^R \\
    \end{array}
  \right)
  \left(
    \begin{array}{cc}
      1 & -1 \\
      1 & 1 \\
    \end{array}
  \right)\\
  =
  \frac{1}{2}
  \left(
    \begin{array}{cc}
      Q^R + Q^A + Q^K & Q^R - Q^A - Q^K \\
      Q^R - Q^A + Q^K & Q^R + Q^A - Q^K \\
    \end{array}
  \right).
\end{split}
\end{equation*}

Furthermore, if we want to express the various pairwise combinations of real-time quantities, for example $Q^< C Q^<$.
We can just substitute $Q^< = (-Q^R + Q^A + Q^K)/2$ into $Q^< C Q^<$ and expand it into nine terms involving $Q^{R/A/K} C Q^{R/A/K}$.

\section{Nine Equations for the generalized NVCs}\label{app:Nine Equations for GNVCs}

The following nine equations are obtained from Eq.\eqref{linear eq for gamma n} in Keldysh's representation with four cases of ${C}^{(i)}$ and ${\Omega}^{(i)}$ (i=1,2,3,4):
\begin{eqnarray}
  \Omega_n^{RR} &=& \avg{t_n^R \bar{G}^R C \bar{G}^R t_n^R} + \sum_{p\neq n}\avg{t_n^R \bar{G}^R \Omega_p^{RR} \bar{G}^R t_n^R},\\
  \Omega_n^{RA} &=& \avg{t_n^R \bar{G}^R C \bar{G}^A t_n^A} +  \sum_{p\neq n}\avg{t_n^R \bar{G}^R \Omega_p^{RA} \bar{G}^A t_n^A},\\
  \Omega_n^{AR} &=& \avg{t_n^A \bar{G}^A C \bar{G}^R t_n^R} + \sum_{p\neq n}\avg{t_n^A \bar{G}^A \Omega_p^{AR} \bar{G}^R t_n^R},\\
  \Omega_n^{AA} &=& \avg{t_n^A \bar{G}^A C \bar{G}^A t_n^A} + \sum_{p\neq n}\avg{t_n^A\bar{G}^A\Omega_p^{AA}\bar{G}^A t_n^A},
\end{eqnarray}
\begin{widetext}
\begin{eqnarray}
  \Omega_n^{RK} &=& \avg{t_n^R \bar{G}^R C \bar{G}^R t_n^K} + \avg{t_n^R \bar{G}^R C \bar{G}^K t_n^A} + \sum_{p\neq n}\Big[ \avg{t_n^R \bar{G}^R \Omega_p^{RK} \bar{G}^A t_n^A} + \avg{t_n^R \bar{G}^R \Omega_p^{RR} \bar{G}^R t_n^K} + \avg{t_n^R \bar{G}^R \Omega_p^{RR}\bar{G}^K t_n^A}\Big],\\
  \Omega_n^{AK} &=& \avg{t_n^A \bar{G}^A C \bar{G}^R t_n^K} + \avg{t_n^A \bar{G}^A C \bar{G}^K t_n^A} + \sum_{p\neq n}\Big[\avg{t_n^A \bar{G}^A \Omega_p^{AK} \bar{G}^A t_n^A} + \avg{t_n^A \bar{G}^A \Omega_p^{AR} \bar{G}^R t_n^K} + \avg{t_n^A \bar{G}^A \Omega_p^{AR} \bar{G}^K t_n^A}\Big],\\
  \Omega_n^{KR} &=& \avg{t_n^K \bar{G}^A C \bar{G}^R t_n^R} + \avg{t_n^R \bar{G}^K C \bar{G}^R t_n^R} + \sum_{p\neq n}\Big[\avg{t_n^R \bar{G}^R \Omega_p^{KR} \bar{G}^R t_n^R} + \avg{t_n^K \bar{G}^A \Omega_p^{AR} \bar{G}^R t_n^R} + \avg{t_n^R \bar{G}^K \Omega_p^{AR} \bar{G}^R t_n^R}\Big],\\
  \Omega_n^{KA} &=& \avg{t_n^K \bar{G}^A C \bar{G}^A t_n^A} + \avg{t_n^R \bar{G}^K C \bar{G}^A t_n^A} + \sum_{p\neq n}\Big[\avg{t_n^R \bar{G}^R \Omega_p^{KA} \bar{G}^A t_n^A} + \avg{t_n^K \bar{G}^A \Omega_p^{AA} \bar{G}^A t_n^A} + \avg{t_n^R \bar{G}^K \Omega_p^{AA}\bar{G}^A t_n^A}\Big],\\
  \Omega_n^{KK} &=& \avg{t_n^K \bar{G}^A C \bar{G}^R t_n^K} + \avg{t_n^K \bar{G}^A C \bar{G}^K t_n^A} + \avg{t_n^R \bar{G}^K C \bar{G}^R t_n^K} + \avg{t_n^R \bar{G}^K C \bar{G}^K t_n^A}\nonumber\\
  &+&\sum_{p\neq n} \Big[ \avg{t_n^R \bar{G}^R \Omega_p^{KK} \bar{G}^A t_n^A} + \avg{t_n^K \bar{G}^A \Omega_p^{AK} \bar{G}^A t_n^A} + \avg{t_n^R \bar{G}^K \Omega_p^{AK} \bar{G}^A t_n^A} + \avg{t_n^R \bar{G}^R \Omega_p^{KR} \bar{G}^R t_n^K}\nonumber\\
  &+&\avg{t_n^K \bar{G}^A \Omega_p^{AR} \bar{G}^R t_n^K} + \avg{t_n^R \bar{G}^K \Omega_p^{AR} \bar{G}^R t_n^K} + \avg{t_n^R \bar{G}^R \Omega_p^{KR} \bar{G}^K t_n^A} + \avg{t_n^K \bar{G}^A \Omega_p^{AR} \bar{G}^K t_n^A} + \avg{t_n^R \bar{G}^K \Omega_p^{AR} \bar{G}^K t_n^A}\Big].
\end{eqnarray}
\end{widetext}

%\bibliographystyle{unsrt}
%\bibliography%{ref}

\end{document}